\documentclass[prl,twocolumn,english,superscriptaddress]{revtex4-1}
\usepackage{amsmath,amssymb,mathrsfs}

\usepackage{amsfonts,bm}
\usepackage{amsmath}
\usepackage{amssymb}
\usepackage{amsthm}
\usepackage{graphicx}
\usepackage{graphics}
\usepackage{subfigure}
\usepackage{color}
\usepackage{bm}
\usepackage{hyperref}
\usepackage{rotating}
\usepackage{comment}

\renewcommand{\>}{\rangle}
\newcommand{\be}{\begin{equation} }
\newcommand{\ee}{\end{equation} }
\newcommand{\ba}{\begin{eqnarray} }
\newcommand{\ea}{\end{eqnarray} }

\newcommand{\down}{\downarrow}

\newcommand{\bpm}{\begin{pmatrix}}
\newcommand{\epm}{\end{pmatrix}}
\newcommand{\bmm}{\begin{matrix}}
\newcommand{\emm}{\end{matrix}}
\newcommand{\up}{\uparrow}

\newcommand{\bea}{\begin{eqnarray}}
\newcommand{\eea}{\end{eqnarray}}
\renewcommand{\v}[1]{\boldsymbol{#1}}

\makeatother

\usepackage{babel}
\begin{document}

\title{Exactly soluble model of boundary degeneracy}
 \author{Sriram Ganeshan}
 \affiliation{Simons Center of Geometry and Physics, Stony Brook, NY 11794}
  \affiliation{Condensed Matter Theory Center, Department of Physics, University of Maryland, College Park, MD 20742, USA}
 \author{Alexey V. Gorshkov}
\affiliation{Joint Quantum Institute, NIST/University of Maryland, College Park, Maryland 20742, USA}
\affiliation{Joint Center for Quantum Information and Computer Science, NIST/University of Maryland, College Park, Maryland 20742, USA}
 \author{Victor Gurarie}
 \affiliation{Physics Department, University of Colorado, Boulder, Colorado 80309, USA}
\author{Victor M. Galitski}
\affiliation{Condensed Matter Theory Center, Department of Physics, University of Maryland, College Park, MD 20742, USA}
\affiliation{Joint Quantum Institute, NIST/University of Maryland, College Park, Maryland 20742, USA}
 \affiliation{School of Physics, Monash University, Melbourne, Victoria 3800, Australia}

\date{\today}

\begin{abstract}
We investigate the topological degeneracy that can be realized in Abelian  fractional quantum spin Hall states with multiply connected gapped boundaries. Such a topological degeneracy (also dubbed as ``boundary degeneracy") does not require superconducting proximity effect and can be created by simply applying a depletion gate to the quantum spin Hall material and using a generic spin-mixing term (e.g., due to backscattering) to gap out the edge modes. We construct an exactly soluble microscopic model manifesting this topological degeneracy and solve it using the recently developed technique [S. Ganeshan and M. Levin, Phys. Rev. B 93, 075118 (2016)]. The corresponding string operators spanning this degeneracy are explicitly calculated. It is argued that the proposed scheme is experimentally reasonable. 
\end{abstract}
\maketitle
There has been significant recent interest and progress in constructing theoretical models that exhibit exotic, non-Abelian anyons  as either  intrinsic excitations or states captured by extrinsic defects in various topological phases~\cite{bravyi1998quantum}. Of particular interest here is the possibility to create such non-Abelian anyons in otherwise Abelian topological states. This was explicitly demonstrated  in theoretical proposals featuring fractional (Abelian) topological states proximity-coupled to superconductors and in bilayer quantum Hall states with extrinsic twist defects~\cite{cheng2012superconducting, barkeshli2012topological, lindner2012fractionalizing, barkeshli2013twist, vaezi2013fractional, clarke2013exotic}. However, there are serious challenges in the experimental realization of these parafermionic models due to a number of poorly-compatible ingredients that have to co-exist in a single system (in particular, superconductivity and topological order). Moreover, in most cases, braiding properties of the non-Abelian anyons are not sufficiently rich to host universal topological quantum computation.

Recent works have shown that multiple gapped boundaries connected with a common topological bulk can play the role of non-Abelian excitations as long as the bulk supports an intrinsic Abelian topological order~\cite{wang2015boundary,lan2015gapped,kapustin2014ground}. The topological ground state degeneracy in these systems has been dubbed as ``boundary degeneracy". In a recent preprint, Barkeshli and Freedman put forward that topological order with a multiply connected gapped boundary can manifest a richer set of topologically protected unitary transformations~\cite{barkeshli2016modular}, raising the possibility of realizing universal quantum computation in systems with no superconducting proximity. 
\begin{figure}[tb]
  \centering
\includegraphics[scale=0.5]{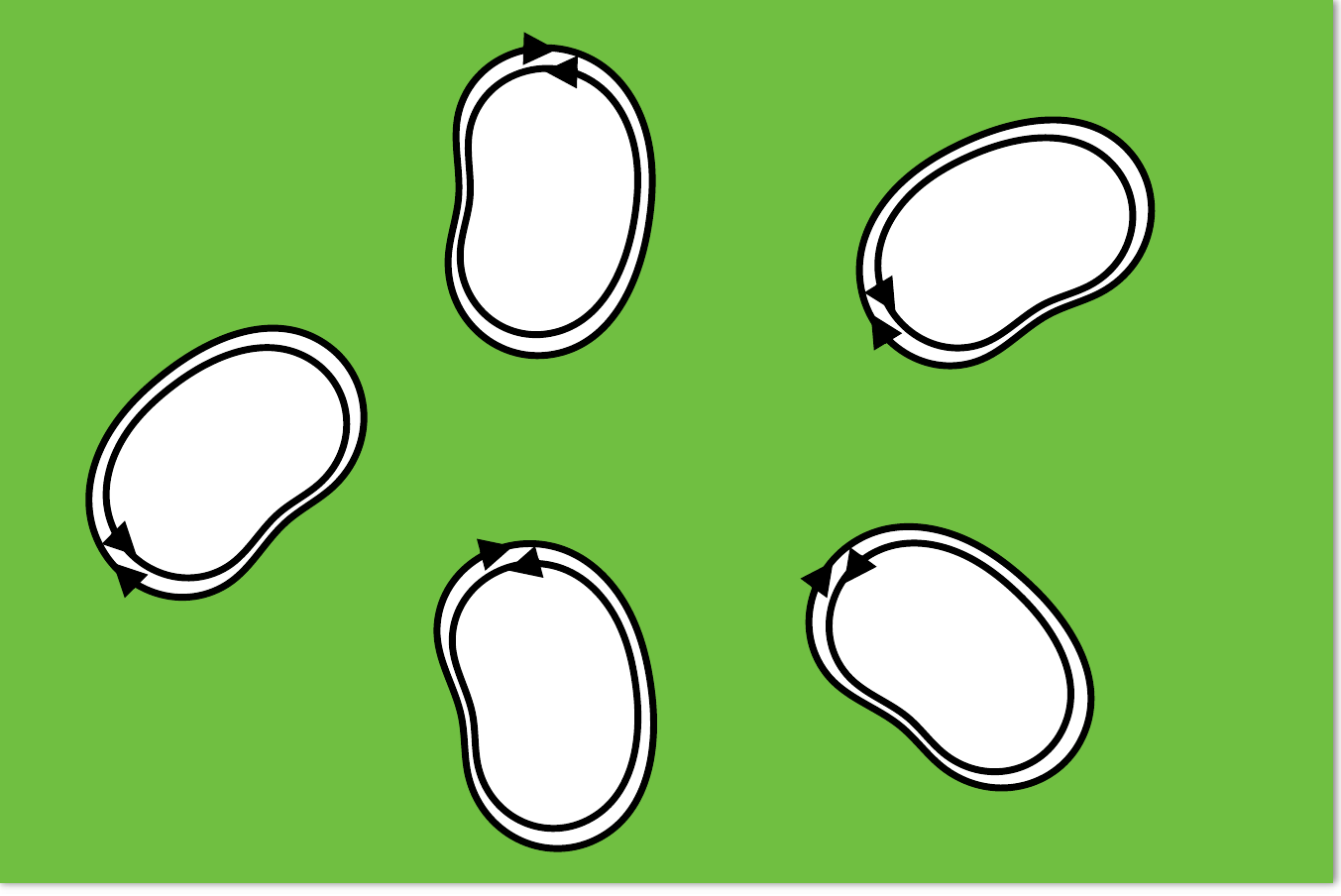}
\caption{FQSH phase (green shading) on a multiply connected 2D surface. Interface between holes (in white) manifest two counter propagating edge modes corresponding to the two spin components.}
     \label{fig:si}
\end{figure}

The simplest system that is a candidate for manifesting boundary degeneracy is a fractional quantum spin Hall (FQSH) state of filling fraction $\nu=1/k$ with multiple holes with a boundary (which can be created using a depletion gate) (Fig.~\ref{fig:si}). Each hole will manifest two counter propagating edge modes corresponding to the two components of spin. We model these edge modes by chiral Luttinger liquids with opposite chiralities. If we allow direct tunneling between the two edge theories, it would gap them out. Punching out $N$ holes and glueing the two spin component together along the edges is equivalent to creating a fractional quantum Hall state on a manifold of genus $N-1$ \cite{wang2015boundary, kapustin2014ground, barkeshli14}, which is known to possess the topological degeneracy $k^{N-1}$.  This proposal for creating topological degeneracy is conceptually simple and could be experimentally implemented immediately when a FQSH is realized. Furthermore, magnetic impurities, which were thought as a nuisance in the current experimental works on QSH effect, can be an advantage towards gapping the edge modes of a FQSH system, which is a necessary step in engineering our topological degeneracy. 

In this work, we construct an exactly soluble microscopic model manifesting topological boundary degeneracy.  Our construction is rooted in the recently developed~\cite{ganeshan2015solution} Hamiltonian formulation. The relevant topological physics manifests in the effective Hilbert space in the non-perturbative backscattering limit. Within this framework, we prove the existence of a robust topological degeneracy and derive the string operators that span this degeneracy. Our approach in this sense differs from the topological quantum field theory methods~\cite{kapustin2014ground} and effective boundary action analysis~\cite{wang2015boundary}. Towards the end, we outline possible experimental platforms to engineer and probe topological degeneracy via multiply connected gapped boundaries.

{\it Model:} We begin with a microscopic model for a perfectly clean homogeneous edge of the $i$th hole modeled by two chiral Luttinger liquids with opposite chiralities, one for each spin direction. We then add non-perturbative backscattering terms that mix the two spin components and gap the edge modes. Finally, we connect all the edges (holes) by a common fractionalized bulk. The formalism we consider naturally incorporates this as a constraint on the allowed charge at the edge. We now construct and systematically solve a microscopic model  that encapsulates all these aspects. 

    The Hamiltonian for a perfectly clean, homogeneous edge of the $i$th hole is given by 
\begin{equation}
H^i_0 = \frac{kv_i}{4\pi}\int_{-L/2}^{L/2}[(\partial_{x}\phi^i_{\up}(x))^{2}+(\partial_{x}\phi^i_{\down}(x))^{2}]dx,
\label{Hclean}
\end{equation}
where $v$ is the velocity of the edge modes of circumference $L$. $\phi^i_{\up/\down}$ are bosonic fields satisfying canonical commutation relations $[\phi^i_{\sigma}(x),\partial_{y}\phi^j_{\sigma'}(y)]=\delta_{ij}\delta_{\sigma \sigma'}\frac{2\pi i}{k_{\sigma}}\delta(x-y)$, where $k_{\uparrow}=-k_{\downarrow}=k$. The density of spin-up electrons at position $y$ at the $i$th hole is given by $\rho^i_\up(y) = \frac{1}{2\pi} \partial_y \phi^i_\up$,  while the density of spin-down electron is $\rho^i_\down(y) = \frac{1}{2\pi} \partial_y \phi^i_\down$. The total charge $Q^i$ and total spin $S^i_z$ on the edge of the $i$th hole are given by $Q^i = Q^i_\up + Q^i_\down$ and $S^i_z = \frac{1}{2}(Q^i_\up - Q^i_\down)$ with
\begin{align*}
Q^i_\sigma = \frac{1}{2\pi} \int_{-L/2}^{L/2} \partial_y \phi^i_\sigma dy, \quad \sigma = \up,\down .
\end{align*}
The spin-up and spin-down electron creation operators at each hole take the form $\psi_\up^{i\dagger} = e^{i k \phi^i_\up}, \quad \psi_\down^{i\dagger} = e^{-i k \phi^i_\down}$.   Note that $H^i$ corresponds to a collection of decoupled edge modes, and the key information that these modes are actually multiply-connected via a common fractionalized bulk is missing. This multiple connectedness of the holes results in two quantization conditions on $Q^i_\up, Q^i_\down$.\\
 
$\quad$ $\quad$ $\quad$ $Q^i_{\up,\down}\in \mathbb{Z}\times 1/k$ $\quad$and$\quad$ $\sum_{i=1}^{N}Q^i_{\up, \down}\in \mathbb{Z}$.\\
 
Physically, these quantization conditions require that the edge modes corresponding to holes contain fractional charges in multiples of $1/k$ and that the net charge on all the holes adds up to be an integer multiple of the electronic charge. For example, the edge of an isolated single hole cannot carry any excess fractional charge.   
    A closely related fact to this quantization is that the bosonic operators $\phi^i_\up(y)$ and $\phi^i_\down(y)$ are actually \emph{compact} degrees of freedom which are only defined modulo $2\pi/k$. Following Ref.~\cite{ganeshan2015solution}, we dynamically impose the quantization on $Q^{i}_\up, Q^{i}_\down$. To this end, we add $ H^i_{lq}=-U\cos(2\pi k Q^i_{\uparrow})-U\cos(2\pi k Q^i_{\downarrow})$ to the edge Hamiltonian of the $i$th hole. 
 We then impose the second condition, corresponding to the global quantization of the total charge on all holes, 
 by adding a global term $H_{gq}=-U\cos(2\pi \sum_i^N Q^i_{\uparrow})-U\cos(2\pi \sum_i^N Q^i_{\downarrow})$. Notice that both quantization conditions are imposed by letting $U\rightarrow \infty$. The Hilbert space corresponding to the clean edge is spanned by the complete orthonormal basis $\{|q^i_\up, q^i_\down, \{n^i_{p\up}\}, \{n^i_{p\down}\}\>\}$, where the quantum numbers $q^i_\up, q^i_\down$ correspond to the total charge associated with the two spin species ranging over $\mathbb{Z}\times1/k$ (subject to $\sum_i q^i_{\up,\down}\in \mathbb{Z}$), while $n^i_{p\up}, n^i_{p\down}$ are the neutral phonon excitations of momentum $p$ ranging over all nonnegative integers for each value of $p = 2\pi/L, 4\pi/L,...$. 
    

 The next step is to add backscattering terms that gap the above defined boundary modes by scattering spin-up electrons to spin-down electrons. A continuum of backscattering terms in a fermionic representation can be expressed as $H^i_{bs}=\int^{L}_{0}\frac{U(x)}{2}\psi^{i\dagger}(x)\psi^i(x)+H.c.$. The corresponding bosonized representation can be written as $H^i_{bs}=\int^L_0 U(x)\cos(k[\phi^i_{\uparrow}(x)+\phi^i_{\downarrow}(x)])$. The total Hamiltonian for the $i$th hole  $H^i_0+H^i_{bs}$ corresponds to a gapped edge in the large $U$ limit. 
\begin{figure}[tb]
  \centering
\includegraphics[scale=0.5]{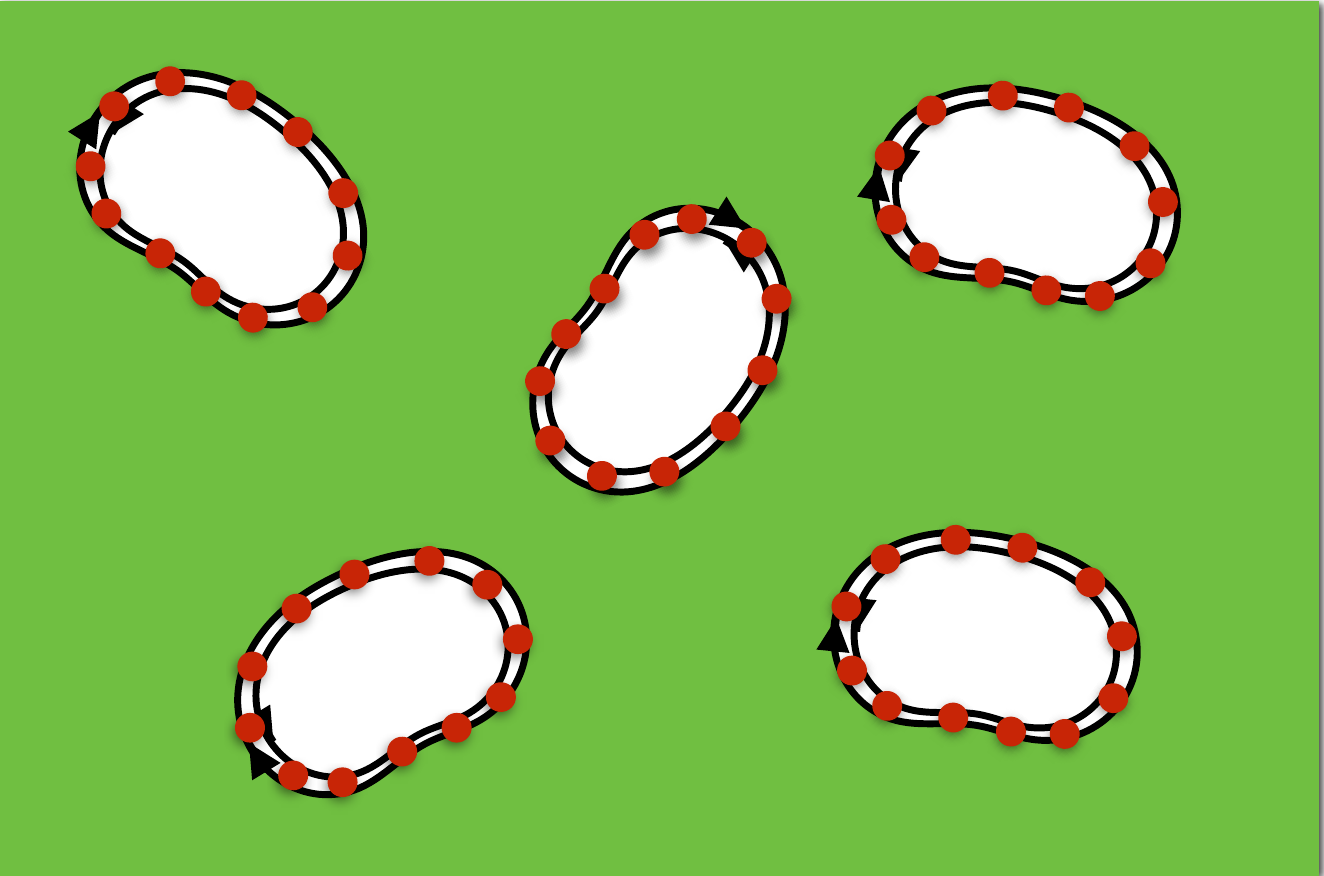}
\caption{FQSH phase (green shading) with gapped boundaries. Red dots denote an array of magnetic impurities that gap edges in the limit of continuum backscattering.}
     \label{fig:holes}
\end{figure}
    Now we are set to write down the full microscopic Hamiltonian corresponding to the $N$ multiply connected hole boundaries: $H=H_{gq}+\sum_{i=1}^N H^i_0+H^i_{bs}+H^i_{lq}$. The Hamiltonian $H$ can be mapped onto a class of exactly soluble Hamiltonians by replacing the  continuum backscattering term $\int^L_0 U(x)\cos(k[\phi^i_{\uparrow}(x)+\phi^i_{\downarrow}(x)])$ with an array of $M$ impurity scatterers $U\sum^M_{j=1}\cos(k[\phi^i_{\uparrow}(x_j)+\phi^i_{\downarrow}(x_j)])$ (see Fig.\ \ref{fig:holes}). The continuum result is then recovered in the thermodynamic limit of $L, M \rightarrow \infty$ with $U$ and $L/M$ fixed. Without loss of generality, we periodically arrange the backscattering terms at each hole as $x_{1..M}=0,..,(M-1)s $, where $s$ is the spacing between two impurity points. After this replacement, the Hamiltonian $H$ is exactly soluble in the limit $U\rightarrow \infty$. To make contact with with the formalism outlined in Ref.\ \cite{ganeshan2015solution}, we rewrite the above model as
   \begin{align}
   H&=H_0-U\sum^{N(M+2)+2}_{i=1}\cos(C_i).
   \label{Ham}
   \end{align}
   In the above notation, the first term  $H_0=\sum^N_i H^i_0$ contains the dynamics of the clean edge. The second term contains the back scattering terms on all the holes and their corresponding charge quantization conditions. We have organized the cosine arguments in the following way. The first $NM$ terms consist of all the back-scattering terms $\{C_{1..M},..,C_{(N-1)M+1...N M}\}=\{k(\phi^1_{\uparrow}(x_{1..M})+\phi^i_{\downarrow}(x_{1..M}))...k(\phi^N_{\uparrow}(x_{1..M})+\phi^N_{\downarrow}(x_{1..M}))\}$. The quantization condition of each hole boundary is given by $\{C_{NM+1},..,C_{NM+N}\}=\{2\pi k Q^1_{\up}...2\pi k Q^N_{\up}\}$ and $\{C_{NM+N+1},..,C_{NM+2N}\}=\{2\pi k Q^1_{\down}...2\pi k Q^N_{\down}\}$. Finally, the two conditions on the total charge are given by $\{C_{N(M+2)+1}, C_{N(M+2)+2}\}=\{2\pi \sum_i^N Q^i_{\uparrow}, 2\pi \sum_i^N Q^i_{\downarrow}\}$.
   
  {\it Degeneracy:} The next step is to calculate the low-energy effective Hamiltonian $H_{\text{eff}}$ and the low-energy Hilbert space $\mathcal{H}_{\text{eff}}$ corresponding to $H$ in the limit $U\rightarrow \infty$. The effective Hamiltonian is given by
\begin{equation}
H_{\text{eff}} =  H_0 - \sum_{i,j=1}^{N(M+2)+2} \frac{(\mathcal{M}^{-1})_{ij}}{2} \cdot \Pi_i \Pi_j,
\label{Heffgenc}
\end{equation}
where the operators $\Pi_1,..., \Pi_{N(M+2)+2}$ are defined by
$\Pi_{i} = \frac{1}{2\pi i} \sum_{j=1}^{N(M+2)+2} \mathcal{M}_{ij} [C_j, H_0]$ and where $\mathcal{M}_{ij}$ is a matrix defined by
$\mathcal{M} = \mathcal{N}^{-1}, \quad \mathcal{N}_{ij} = -\frac{1}{4\pi^2}[C_i,[C_j,H_0]]$. $\Pi_i$ operators satisfy $[C_i, \Pi_j] = 2\pi i \delta_{ij}$ by construction. The simple physical intuition is that the low-energy physics of $H$ in the limit $U\rightarrow \infty$ does not contain the dynamics of $C_i$'s. Thus the term generating the dynamics $\frac{(\mathcal{M}^{-1})_{ij}}{2}\Pi_i \Pi_j$ must be removed from the effective Hamiltonian. This effective Hamiltonian is defined on an effective Hilbert space $\mathcal{H}_{\text{eff}}$, which is a \emph{subspace} of the original Hilbert space $\mathcal{H}$ and which consists of all states $|\psi\>$ satisfying $\cos(C_i)|\psi\> = |\psi\>, \quad i=1,...,N(M+2)+2$.  We can directly find the creation and annihilation operators for $H_{\text{eff}}$ by finding all operators $a$ that obey $[a, H_{\text{eff}}]=E a$. Putting this all together, we see that the most general possible creation/annihilation operator for $H_{\text{eff}}$ is given by
\begin{align*}
a_{ipm} = \sqrt{\frac{k}{4\pi|p|s}} \int_{-L/2}^{L/2} &[(e^{ipy}\partial_y \phi^i_\up + e^{2ip x_m} e^{-ipy} \partial_y \phi^i_\down). \\
         & \cdot \Theta(x_{m-1}<y<x_{m})]dy
\end{align*}
Here the index $m$ runs over $m=1,...,M$, $i$ runs over the holes $i = 1,...,N$, while $p$ takes values $\pm \pi/s, \pm 2\pi/s,...$. The operators are normalized to yield $[a_{ipm}, a_{i'p'm'}^{\dagger}] = \delta_{pp'} \delta_{mm'}\delta_{ii'}$ for $p,p' > 0$. The cosine terms imposing quantization/compactness condition naturally forbids $a$ to be an explicit function of the bosonic field $\phi$.

We now construct a complete set of commuting operators for labeling the eigenstates of $H_{\text{eff}}$. In order to do this, we consider the integer and skew-symmetric $(N(M+2)+2) \times (N(M+2)+2)$ matrix $\mathcal{Z}_{ij}$ defined by $\mathcal{Z}_{ij} = \frac{1}{2\pi i} [C_i, C_j]$. Let $C_i' = \sum_{j=1}^{N(M+2)+2} \mathcal{V}_{ij} C_j + \chi_i$ for some matrix $\mathcal{V}$ such that $[C_i', C_j'] = 2\pi i \mathcal{Z}_{ij}'$, where $\mathcal{Z}' = \mathcal{V} \mathcal{Z} \mathcal{V}^T$. The offset $\chi_i$ must be chosen to be $\chi_i = \pi \cdot \sum_{j < k} \mathcal{V}_{ij} \mathcal{V}_{ik} \mathcal{Z}_{jk} \pmod{2\pi}$ such that $e^{iC_i'}|\psi\> = |\psi\>$ is satisfied
for any $|\psi\> \in \mathcal{H}_{\text{eff}}$. We then find a matrix $\mathcal{V}$ with integer entries and determinant $\pm 1$, such that $\mathcal{Z}'$ takes the simple form
\begin{equation}
\mathcal{Z}' = \bpm 0_N & -\mathcal{D}_N & 0 \\
          \mathcal{D}_N & 0_N & 0 \\
          0 & 0 & 0_{N M+2} \epm, \quad \mathcal{D}_N = \bpm 1 & 0 & \dots & 0 \\
                                                  0 & k & \dots & 0 \\
                                                  \vdots & \vdots & \vdots & \vdots \\
                                                   0 & 0 & \dots & k \epm.
\label{zprime}
\end{equation}
Here $0_{N}$ denotes an $N \times N$ matrix of zeros. $\mathcal{V}$ is an integer change of basis that puts $\mathcal{Z}$ into \emph{skew-normal} form. In the $\v C'$ basis, the diagonalized low-energy effective Hamiltonian $H_{\text{eff}}$ takes the form
\begin{equation}
H_{\text{eff}} =\sum_{i=1}^{N}\sum_{m=1}^{M} \sum_{p > 0} v p a_{ipm}^\dagger a_{ipm} + F(C_{2N+1}',..., C_{N(M+2)+2}'),
\label{Hefffm}
\end{equation}
where the sum runs over $p = \pi/s, 2\pi/s,...$ and
where $F$ is some quadratic function of $NM+2$ variables associated with the $0_{NM+2}$ block of the $\mathcal{Z}_{ij}'$ matrix. The exact form of $F$ does not play a role in the analysis to follow and we keep it general (even though it can be computed following Ref.~\cite{ganeshan2015solution}). 
Using the commutation algebra of the $C_i'$ operators, we can construct the complete set of operators that commute with each other and with $H_{\text{eff}}$.
The effective Hilbert space $\mathcal{H}_{\text{eff}}$ is then spanned by
the unique simultaneous eigenstates $\{|\v{\alpha},\v{q},\{n_{ipm}\}\>\}$  satisfying
\begin{align}
&e^{iC_{1,N+1}'} |\v{\alpha},\v{q},\{n_{ipm}\}\> = |\v{\alpha},\v{q},\{n_{ipm}\}\>,\nonumber\\
&e^{iC_{2,..,N}'/k} |\v{\alpha},\v{q},\{n_{ipm}\}\> = e^{i2\pi \alpha_{2..N}/k}|\v{\alpha},\v{q},\{n_{ipm}\}\>,\nonumber \\
&e^{iC_{N+2,..,2N}'} |\v{\alpha},\v{q},\{n_{ipm}\}\> = |\v{\alpha},\v{q},\{n_{ipm}\}\>,\nonumber \\
&C_{2N+1...N(M+2)+2}' |\v{\alpha},\v{q},\{n_{ipm}\}\> = 2\pi q_{1..NM+2} |\v{\alpha},\v{q},\{n_{ipm}\}\>,\nonumber \\ 
&a_{ipm}^\dagger a_{ipm} |\v{\alpha},\v{q},\{n_{ipm}\}\> = n_{ipm} |\v{\alpha},\v{q},\{n_{ipm}\}\>. \label{quantnumdef3}
\end{align}
Here the label $n_{ipm}$ runs over non-negative integers, while $\v{\alpha}$ is an abbreviation for the $(N\!-\!1)$-component integer vector $(\alpha_2,...,\alpha_{N})$ where $\alpha_{2..N}$'s run over $\{0...k-1\}$.
$\{|\v{\alpha},\v{q},\{n_{pm}\}\>\}$ basis states are also eigenstates of $H_{\text{eff}}$ with the total energy given by
\begin{equation}
E = \sum_{i=1}^N\sum_{m=1}^M \sum_{p > 0} v p n_{ipm}+ F(2\pi q_{1},...,2\pi q_{N M+2}).
\label{Efmsc}
\end{equation}
There are two important features of the above spectrum. a) $E$  has a finite \emph{energy gap} of order $v/s$ where $s = L/M$. b) The spectrum  $E$ is independent of the quantum numbers $\v{\alpha}$. In other words, every state, including the ground state, has a degeneracy of
\begin{equation}
D = k^{N-1}
\label{degfmsc}
\end{equation}
since this is the number of different values that $\v{\alpha}$ ranges over. This degeneracy agrees with the prediction made in the introduction.

{\it String Operators:} In the above analysis, we were able to identify quantum numbers and the complete set of commuting operators associated with the effective Hilbert space. From these commuting operators we can deduce the so-called ``string operators" that span the degenerate subspace. The string operators in the primed basis are given by $\{e^{iC_{2,..,N}'/k}, e^{iC_{N+2,..,2N}'/k}\}$. In the unprimed basis, these operators are defined as 
\begin{align}
\{e^{i2\pi (Q^i_{\up}-Q^i_{\down})},\prod^{j}_{r=1} e^{i(\phi_{\up}^r(x)+\phi_{\down}^r(x))}\}\nonumber, \\ i=1,..,N-1,\  j=i+1,..,	
\label{stringops}
\end{align}
Note that the above operators are closely related to the parafermion operators and are fixed by the non-unique choice of $\mathcal{V}$. One can obtain the matrix representation of these string operators by acting in the basis states spanned by the degenerate ground state subspace $|\v{\alpha},0,0\>\equiv |\v{\alpha}\>$. 
\begin{align}
	e^{\pm iC_{i}'/k}|\v{\alpha}\>&=e^{\pm i2\pi \alpha_i/k}|\v{\alpha}\>\nonumber\\
	e^{\pm iC_{i+N}'/k}|\v{\alpha}\>&=|\v{\alpha}\pm \v{e_{i-1}}\>,\,\,i=2...N. \nonumber
\end{align}
Here $\v{e_i}$ denotes the $(N\!-\!1)$-component vector $\v{e_i} = (0,...,1,...,0)$ with a ``1'' in the $i$th entry and $0$ everywhere else. The addition of $\v{e_i}$ is performed modulo $k$. Note that the above equations imply that the operators $e^{\pm i C_i'/k}$ act like ``clock'' matrices for $i=2,...,N$, while the operators $e^{\pm i C_i'/k}$ act like ``shift'' matrices for $i=N+2,...,2N$; thus these operators generate a generalized Pauli algebra (a.k.a $\sigma_z, \sigma_x$).

{\it Topological Robustness:} 
 Having established the ground state degeneracy in the $U\rightarrow \infty$ case of our toy model, we proceed to describe finite-$U$ corrections to $H_{\text{eff}}$. Notice that we only seek finite-$U$ corrections to the back-scattering terms that gap the edge. 
 In other words, consider Eq.~(\ref{Ham}) to be of the form $H = H_0 - U \sum^{NM}_{i=1}\cos(C_{i}) - U'\sum^{N(M+2)+2}_{i=NM+1} \cos(C_i)$ in the limit where $U$ is finite but $U' \rightarrow \infty$ ($U'$ are associated with the quantization condition). In this case, the finite-$U$ corrections only generate (instanton-like) tunneling processes of the form $C_{i} \rightarrow C_{i} - 2\pi n_i$ (for $i=1...NM$).
 
 The thermodynamic limit we consider is where $L, M \rightarrow \infty$ with $U$ and $L/M$ fixed. Notice that the boundary corresponding to each hole has a finite \emph{energy gap} in this limit (of order $v/s$, where $s = L/M$). Due to the gapped spectrum, we can employ perturbative methods to probe the degeneracy. The most general low-energy operator generating finite-$U$ corrections to the ground state can be written as $e^{i \sum_{j=1}^{NM} m_j \Pi_j} \cdot \epsilon_{\v{m}}$ with the sum running over the $N M$-component integer vectors $\v{m} = (m_1,...,m_{NM})$~\cite{ganeshan2015solution}. Here, the $\epsilon_{\v{m}}$ are unknown functions of $\{a_{ipm}, a_{ipm}^\dagger, C_{2N+1,...N(M+2)+2}'\}$ that vanish in the limit $U\rightarrow \infty$. The $\Pi_i$ operators are conjugate to the $C_i$'s ($[C_i, \Pi_j]=2\pi i \delta_{ij}$) and thereby generate   tunneling events associated with the finite-$U$ corrections.
 Since the spectrum is gapped in the limit of interest, the ground state degeneracy and the gap are robust against small perturbations. The lowest-order non-vanishing matrix elements splitting the degeneracy within the ground state come from the simultaneous single-instanton tunneling event at all $M$ impurity points of a given hole ($m_1=...=m_M=1$, which is an $M$th-order instanton process). This lowest-order splitting is suppressed by a factor of $\sim e^{-\text{const}.M \sqrt{U}}$~\cite{coleman1988aspects,ganeshan2015solution}, which vanishes in the thermodynamic limit of $M\rightarrow \infty$, exemplifying the topological nature of the degeneracy.
\begin{figure}[htb!]
  \centering
\includegraphics[width=9.3cm,height=3.8cm]{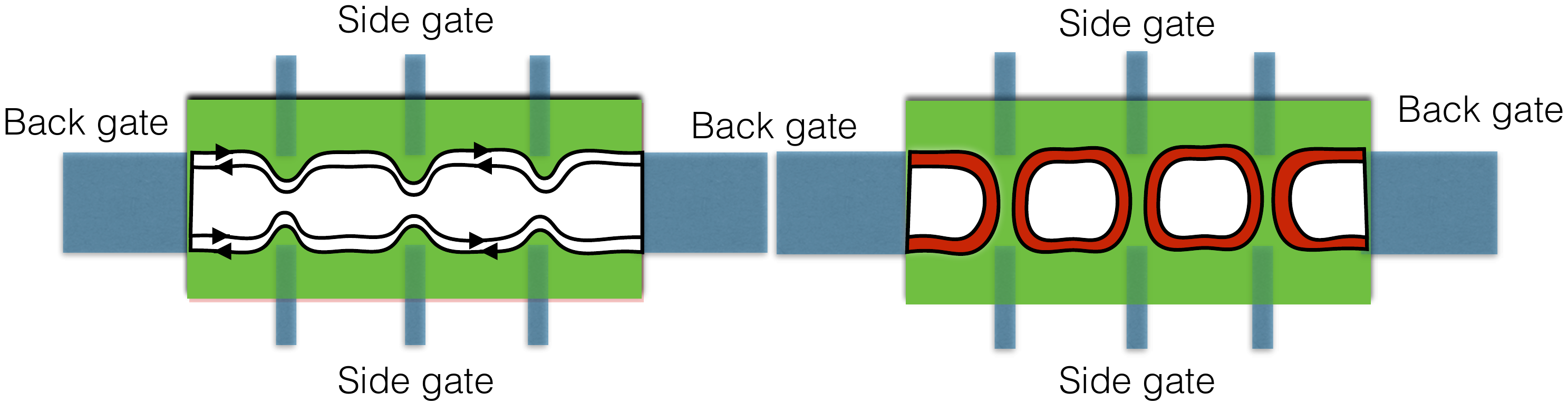}\\
 \quad\quad\text{Electron tunneling}   \quad \quad    \text{\quad \quad Quasiparticle tunneling}
        \caption{Schematic to create topological degeneracy (top view): (Left) FQSH (green shading) with an elongated depletion region (white region) controlled by a back gate. The side gates create QPC that weakly scatters electrons across the trench. (Right) FQSH with doped magnetic impurities that gap the edge. Each hole is shown in white with a red shading denoting a gapped boundary. The side gate voltage is tuned to the strong back scattering limit or quasiparticle tunneling regime. The side gates allow exchange of quasiparticles between the disconnected gapped boundaries.}
        \label{fig:expholes}
\end{figure}

{\it Experimental realization:} The proposed model for topological degeneracy can be realized in a variety of systems where edges around punctures of a conjugate pair of Abelian fraction quantum Hall states can be gapped via backscattering. First, an electron-hole bilayer can exhibit the desired pair of conjugate Abelian fractional quantum Hall states, while top and bottom gates can be used to puncture holes, whose edges can be coupled via electron tunneling \cite{barkeshli14}. Second, a back gate in an electronic FQSH system can be used to puncture holes, while magnetic impurities can be used to flip the spin and thus couple the edges. In Fig~\ref{fig:expholes}, we outline a generalization of an architecture that has been used in fractional quantum Hall experiments~\cite{goldman1995resonant}. The idea is to create a central depletion region using 
a back gate. The side gates create a quantum point contact that can pinch off the trench and create multiply connected regions in the topological state. Notice that in the dual limit after the pinch-off the holes exchange fractional quasiparticles thereby changing the topological sectors controlled by the side gate.

Third, ultracold dipoles, such as magnetic atoms \cite{paz13,baier15}, polar molecules \cite{yan13,yao13c}, and Rydberg atoms \cite{schaus15,barredo15}, pinned in optical lattices can be used to realize spin models whose ground states correspond to bilayer fractional quantum Hall states \cite{yao15}. It is possible that the ground state of such a bilayer system can be tuned to the desired conjugate pair of Abelian fractional quantum Hall states, in which case focused laser beams can be used to locally modify the spin model to effectively puncture holes and couple the resulting edges. Fourth, with the help of synthetic gauge fields and contact interactions, two internal states of ultracold atoms can exhibit the  FQSH effect, while focused laser beams can be used to puncture holes and induce transitions between the two internal states, thus coupling the edges \cite{liu16}. Finally, photonic implementations in radio-frequency \cite{ningyuan15}, microwave \cite{hafezi14b}, and optical \cite{cho08b,umucalilar12,umucalilar13,hayward12,hafezi13c,maghrebi15} domains can also be envisioned.

\begin{acknowledgments}
{\it Acknowledgements}:  SG gratefully acknowledges Michael Levin for an earlier collaboration, where this method was developed, and several useful discussions on the current project. SG was supported by LPS-MPO-CMTC, JQI-NSF-PFC, Microsoft Station Q. VMG was supported by DOE-BES (DESC0001911) and the Simons Foundation. VG was supported by DMR-1205303 and PHY-1211914. VG would also like to thank hospitality of School of Physics, Monash University where this work was conceived. AVG thanks Fangli Liu for discussions and acknowledges support by AFOSR, NSF PIF, ARO, ARL, and NSF PFC at the JQI. After the completion of this paper, we learned of related work \cite{barkeshli16} studying topological degeneracy and non-Abelian braiding in electron-hole bilayers with boundaries proximity-coupled to a superconductor.
\end{acknowledgments}

\bibliographystyle{my-refs}
\bibliography{references1.bib}
\end{document}